\documentclass[fleqn,usenatbib]{mnras}

\usepackage{newtxtext,newtxmath}

\usepackage[T1]{fontenc}

\DeclareRobustCommand{\VAN}[3]{#2}
\let\VANthebibliography\thebibliography
\def\thebibliography{\DeclareRobustCommand{\VAN}[3]{##3}\VANthebibliography}

\usepackage{graphicx}	
\usepackage{amsmath}	

\title[PSR J0437-4715]{Off-Pulse GeV Emission from PSR J0437–4715 and Statistical Properties of Bow-Shock Pulsar Wind Nebulae}

\author[Ou]{
Ziwei Ou,$^{1,2}$\thanks{E-mail: ziwei@sjtu.edu.cn}
\\
$^{1}$Tsung-Dao Lee Institute, Shanghai Jiao Tong University, Shanghai 201210, China\\
$^{2}$State Key Laboratory of Dark Matter Physics, Shanghai Jiao Tong University, Shanghai, 200240, China
}

\date{Accepted XXX. Received YYY; in original form ZZZ}

\pubyear{\the\year{}}

\newcommand{\g}{$\gamma$-ray}
\newcommand{\lat}{\textit{Fermi}-LAT}

\begin{document}
\label{firstpage}
\pagerange{\pageref{firstpage}--\pageref{lastpage}}
\maketitle

\begin{abstract}
PSR~J0437$-$4715 is a $\gamma$-ray millisecond pulsar, which has been detected by \textit{Fermi} Large Area Telescope (LAT). To understand the nature, we analyze the GeV $\gamma$-ray data obtained with \textit{Fermi}-LAT around the pulsar region. Based on the pulsar timing ephemeris, we derived the $\gamma$-ray pulse profile and defined on-pulse and off-pulse phase intervals. A binned likelihood analysis was performed to investigate the spectral properties of the pulsar across different phase ranges. No spatial extension was detected for off-pulse, with an upper limit radius of 0.12$^{\circ}$. The relationship between $L_{\gamma}$ and $L_{X}$ is found as $L_{\gamma} \sim L_{X}^{0.6}$. The conversion efficiency from spin-down power to GeV emission of the outer gap model is consistent with a termination shock located close to the pulsar. We discuss the potential nature of off-pulse GeV emission and the connection to bow shock pulsar wind nebulae (PWNe). As a benchmark, the detailed phase-resolved properties of PSR~J0437$-$4715 are first examined; we then place this pulsar within a larger sample of 17 bow-shock PWNe to test whether its emission characteristics conform to the global scaling relations between $\gamma$-ray luminosity, X-ray luminosity, and bow-shock radius. These two  approaches allow us to assess the origin of the off-pulse emission and the universality of particle acceleration physics in MSP nebulae.
\end{abstract}



\section{Introduction} \label{sect:intro}

When a pulsar moves through the interstellar medium (ISM) at a speed exceeding the ISM sound speed, a bow shock pulsar wind nebula (PWN) is expected to form \citep{Aldcroft1992,Bucciantini2005}. The shape of the shock and the appearance of the nebula may depend on the angle between the pulsar's velocity vector and its spin axis \citep{Vigelius2007}. Additionally, inhomogeneities in the ambient medium can also influence the nebula's morphology \citep{Toropina2019}. The ISM matter is compressed and heated as it passes through the forward shock, which can lead to excitation of ISM atoms followed by radiative de-excitation in the shocked ISM \citep{Bucciantini2001,Nikonorov2025}. Strong emission in the Lyman and Balmer lines is expected due to collisional excitation. Since emission in the Lyman lines is strongly absorbed by the ISM, bow shocks are most easily detected in H$\alpha$ \citep{Bucciantini2002,Brownsberger2014}. 

Several pulsars exhibit elongated or tail-like PWN morphologies in X-rays and radio, commonly attributed to synchrotron radiation from shocked pulsar winds \citep{Moon2004,Gaensler2004,Gaensler2006b,Kim2020}. These structures serve as critical tracers of the interaction between the relativistic pulsar wind and the ambient medium. Such extended features are found to be aligned with the pulsar's proper motion direction, which support their association with bow shock PWN \citep{Kargaltsev2017}. In the X-ray band, these emissions arise from highly relativistic electrons and positrons accelerated at the termination shock of the pulsar wind \citep{Auchettl2015}. In bow shock PWN, the anisotropic pressure of the ambient medium confines the shocked flow into a cometary tail extending behind the pulsar \citep{Marelli2019}. The X-ray morphology may have a feature of compact emission near the pulsar \citep{Kargaltsev2007}. 

While both young pulsars and millisecond pulsars (MSPs) can in principle produce bow shock nebula, their manifestations differ markedly due to distinct evolutionary stages and environments \citep{Stappers2003,Gaensler2006a}. Young pulsars are typically embedded within their supernova remnants, where the surrounding medium is relatively static, often resulting in approximately spherical PWN morphologies during their early life. In contrast, MSPs are ancient, recycled objects that have long escaped their birth sites and now traverse the general ISM at high velocities \citep{Gaensler2002}. Their modest spin-down luminosities ($\sim 10^{33}-10^{34}$~erg~s$^{-1}$) make their pulsar winds more susceptible to confinement in the ISM, which produce compact, bow shock or cometary PWN morphologies. However, bow shock nebulae are not the only possible outcome for MSPs. In cases of low space velocity or unusually dense environments, more isotropic morphologies could in principle emerge \citep{Bucciantini2001}. Nevertheless, the MSP nebula is expected to be compact and often point-like at GeV energies, as the termination shock remains close to the pulsar \citep{Kirk2009,Bykov2017}.

PSR~J0437$-$4715 is a MSP with spin period of 5.8 ms and spin-down energy loss of $\dot{E} \sim 6\times10^{33}\ \rm erg/s$ \citep{Johnston1993}. The binary period of the system is about 5.74 days with a cool white dwarf as the pulsar's companion \citep{Reardon2024}. It is located at a precisely measured distance of $156.79 \pm 0.25\ \rm pc$, which makes it the closest known pulsar. The faint extended X-ray emission detected in the vicinity of the pulsar is associated with the bow shock PWN \citep{Rangelov2016}. It has been shown that accelerated leptons from the nebula of PSR~J0437$-$4715 can be responsible for the enhancement of the positron fraction above a few GeV \citep{Bykov2019}.

We conducted data analysis of \lat\ for PSR~J0437$-$4715 in Section~\ref{sec:fermi-data} together with the results. The discussion is provided in Section~\ref{sec:orgin}. Since the cutoff energy is important for \g\ MSP, phased-resolved cutoff energy for each bin and their relation with \g\ luminosity are presented (Section~\ref{sec:magnetosphere}). The conversion efficiency is obtained to discuss the energy converted from pulsar to \g\ emissions (Section~\ref{sec:efficiency}). Since PSR~J0437$–$4715 is one of the few MSPs with a well-measured bow-shock PWN in X-rays \citep{Rangelov2016}, it serves as an ideal testbed for linking individual nebular properties to population-wide trends. In Section~\ref{sec:population}, we investigate a sample of bow-shock PWN with detected X-ray and \g\ . We therefore expand our scope to a compiled sample of 17 \g\ pulsars with detected bow-shock nebulae, investigating the empirical correlations among $L_{\gamma}$, $L_X$, and $R_{\rm bs}$. By comparing the position of PSR~J0437$-$4715 in these correlation diagrams with the best-fit relations derived from the full sample, we can evaluate whether this nearby MSP is a typical representative of the bow-shock population or an outlier that demands a distinct physical interpretation.

\section{\lat\ Data Analysis} \label{sec:fermi-data} 

The \lat\ data utilized in this study spans from August 8, 2008 to November 19, 2025. We selected all events within the 0.3-30 GeV energy range and a region-of-interest of $15^{\circ}\times15^{\circ}$ for PSR~J0437$-$4715. The zenith angles of events larger than 90° are excluded to reduce the contamination from Earth's limb. We selected good quality front and back-converted events in the dataset (i.e. \texttt{evclass==128, evtype==3}). In \lat\ data analysis, the \texttt{evclass} defines the overall purity threshold (background rejection level) for a photon dataset, while the \texttt{evtype} subdivides events within a given class based on their reconstructed quality characteristics, such as conversion layer, angular resolution, or energy dispersion. Its coordinates (R.A.=69.316$^{\circ}$, Dec.=-47.253$^{\circ}$) are given by the DR4 of 4FGL catalog. The analysis was performed using the Fermi Science Tools \textit{v11r5p3} and Fermipy \textit{1.2.0} packages. The latest \lat\ source catalog 4FGL-DR4 was used to construct a source model. For the diffuse background components, we adopt the Galactic diffuse emission model gll\_iem\_v07 and isotropic extragalactic emission model P8\_R3\_V3.

\subsection{Timing Analysis}

\begin{figure*}
    \centering
    \includegraphics[width=0.75\linewidth]{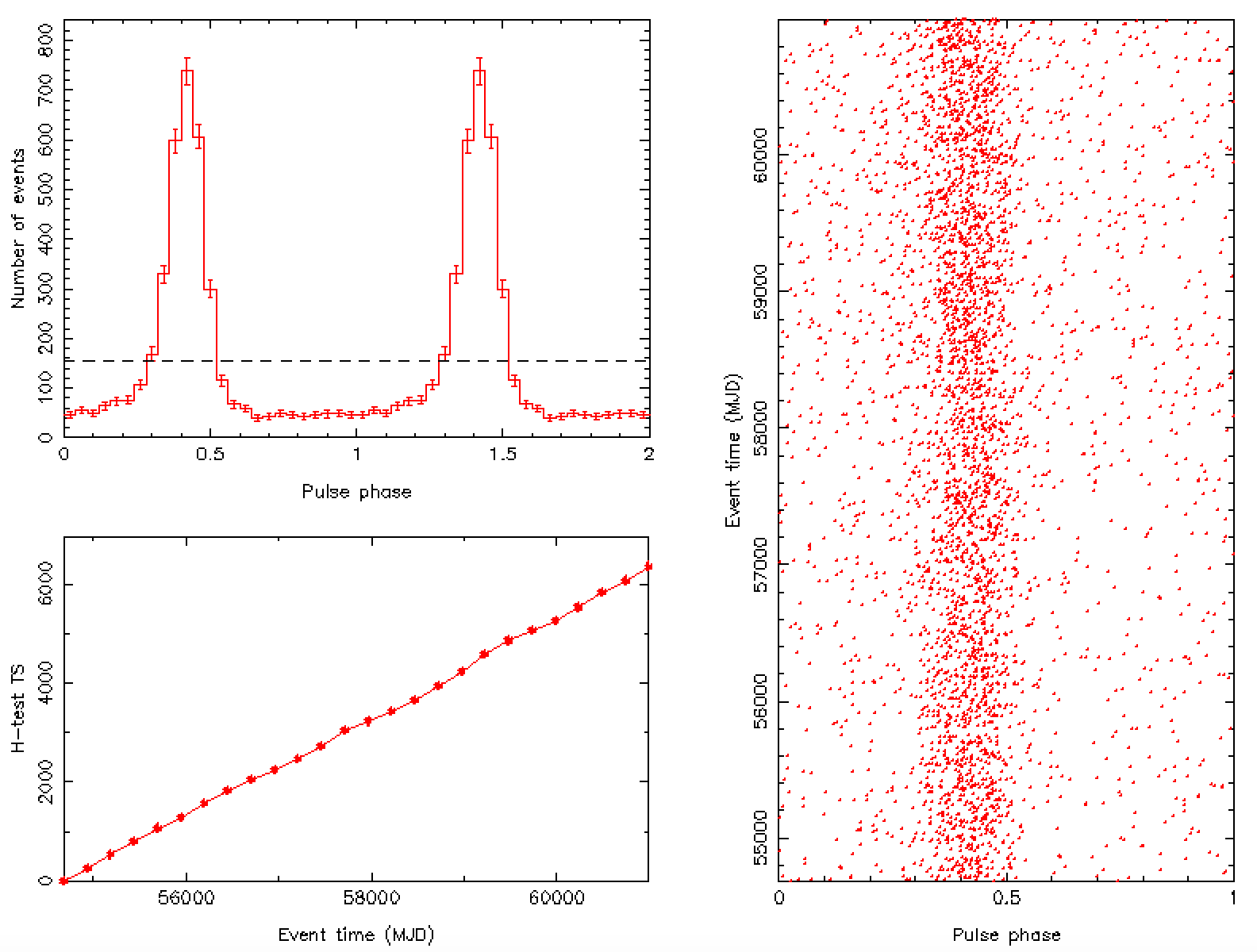}
    \caption{PSR~J0437$-$4715 timing results from \textit{Tempo2} with the \textit{Fermi} plug-in. Top-left panel: phase histogram of the analyzed \lat\ data. Two full rotational phase are shown here. Bottom-left panel: H-test significance (TS) as a function of time. Right panel: pulse phase for each \g\ event vs. time.}
    \label{fig:profile}
\end{figure*}

PSR~J0437$-$4715 is a bright pulsar in the Third Fermi Pulsar Catalog (3PC) \citep{Smith2023}. It is the counterpart of the \g\ source 4FGL~J0437.2$-$4715 in 4FGL catalog. To obtain the off-pulse \g\ emission of the pulsar, we selected photons in a radius of $0.6^{\circ}$ to perform \textit{H}-test statistics. We adopted the timing ephemeris from 3PC \citep{Smith2023}, which are shown in Table~\ref{tab:ephemeris}. \textit{Tempo2} \citep{Hobbs2006} with \textit{Fermi} plug-in were used to produce \g\ pulse profile \citep{Ray2011}. The pulse profile together with the H-test results are shown in Fig.~\ref{fig:profile}. We were able to obtain Fourier template profiles for the whole time span, generate the times of arrival (TOAs), and obtain timing solutions by fitting the TOAs with frequency derivatives. 

\begin{figure*}
    \centering
    \includegraphics[width=0.95\linewidth]{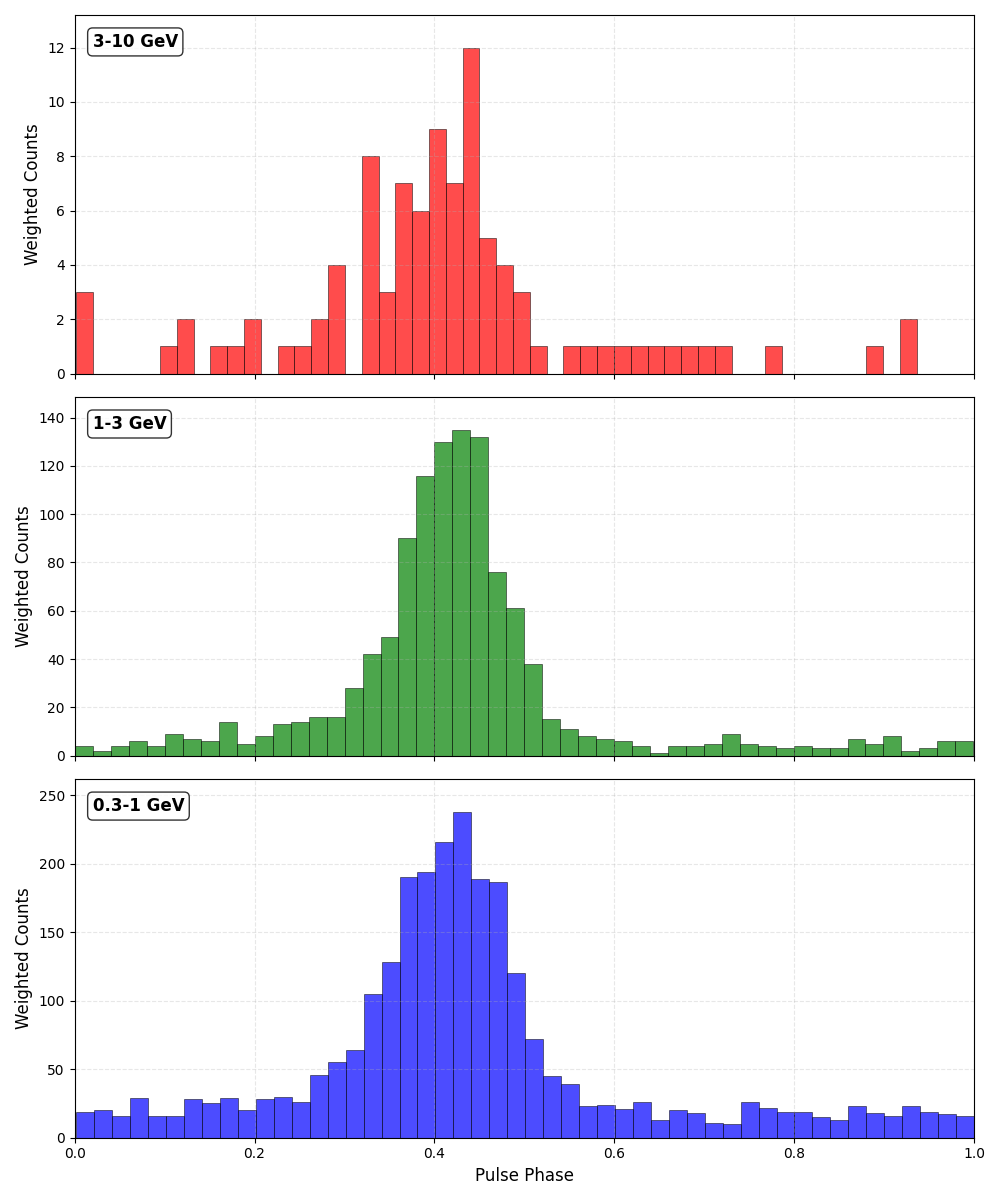}
    \caption{Weighted pulse profile of PSR~J0437$-$4715 at different energies.}
    \label{fig:weight}
\end{figure*}

We also investigate the weighted pulse of PSR~J0437$-$4715 using \texttt{gtsrcprob}. Figure~\ref{fig:weight} shows weighted pulse profile of PSR~J0437$-$4715 at different energies. No significant discrepancy of pulse profile is found in different energy bands. Based on the pulse profile of the \g\ pulsar, we defined phase 0.28–0.52 as the on-pulse phase range and phases 0–0.24 pulse 0.56–1.0 as the off-pulse phase range.

\begin{table}
    \centering
    \begin{tabular}{ll}
    \hline
    Parameter & Value \\\hline
       Pulsar Name  & PSR~J0437$-$4715 \\
       R.A.  & 04:37:15.896 \\
       Dec.  & -47:15:09.11 \\
       MJD range & 55619-57968 \\
       Pulse frequency $\rm s^{-1}$ & 173.6879458 \\
       First derivative of frequency $\rm s^{-2}$ & $-1.72836\times10^{-15}$\\
       Epoch of frequency (MJD) & 54500 \\
    \hline
    \end{tabular}
    \caption{Timing Ephemeris of PSR~J0437$-$4715 taken from 3PC.}
    \label{tab:ephemeris}
\end{table}

\begin{table*}
    \centering
    \begin{tabular}{lccccc}
    \hline
        Phase Range & Flux & $\Gamma$ & $d$ & TS & log(L) \\
         & ($\rm ph\ s^{-1}\ cm^{-2}$) & & & & \\\hline
        whole & $(1.04 \pm 0.02) \times 10^{-08}$ & $1.82 \pm 0.05$ & $0.82 \pm 0.05$ & 7029 & -141844.2517\\\hline
        on-pulse & $(9.72 \pm 0.17) \times 10^{-09}$ & $1.81 \pm 0.04$ & $0.81 \pm 0.05$ & 1403 & -154129.9931\\\hline
        off-pulse (PL) & $(1.09 \pm 0.11) \times 10^{-09}$ & $2.71 \pm 0.09$ & - & 210 & -269178.3188\\
        off-pulse (PLSC) & $(1.11 \pm 0.09) \times 10^{-09}$ & $1.76 \pm 0.06$ & $0.83 \pm 0.04$ & 207 & -268386.2251 \\
    \hline
    \end{tabular}
    \caption{Binned likelihood analysis results for whole data, on-pulse and off-pulse of PSR~J0437$-$4715. These include flux, photon index $\Gamma$, distance, TS values, and log-likelihood values.}
    \label{tab:fit-result}
\end{table*}

\subsection{Likelihood Analysis}

We perform binned likelihood analysis to the \lat\ data. There are 17 sources within $5^{\circ}$ from PSR~J0437$-$4715 in the 4FGL catalog. The spectral parameters in the source models of them were set free. In addition, the parameters of Galactic diffuse emission and isotropic extragalactic emissions were also set free. 

For the off-pulse of pulsar, we first consider a simple power-law (PL): 

\begin{equation}
    \frac{dN}{dE} = N_0 (\frac{E}{E_0})^{-\Gamma}
\end{equation}

where $N_0$, $E_0$ and $\Gamma$ are normalization, pivot energy and photon index, respectively. For the emissions from pulsar, either the whole spin phase or on-pulse phase, an exponential cutoff power-law model (PLSC) is adopted to fit the data: 

\begin{equation} \label{eq:plsc}
    \frac{dN}{dE} = N_0 \left( \frac{E}{E_0} \right)^{-\Gamma-\frac{d}{2}\ln\frac{E}{E_0} - \frac{db}{6} \ln^2\frac{E}{E_0} - \frac{db^2}{24} \ln^3\frac{E}{E_0}}
\end{equation}

where $\Gamma$ $d$ and $b$ are the photon index, the local curvature at $E_0$, and the parameter described shape of the exponential cutoff, respectively. We fixed $b=2/3$ as provided by \lat\ pulsar catalogs. We obtain a TS value of 210 for off-pulse (Table~\ref{tab:fit-result}).

\subsection{Spatial Analysis} \label{subsec:spatial}

\begin{figure*}
    \centering
    \includegraphics[width=0.75\linewidth]{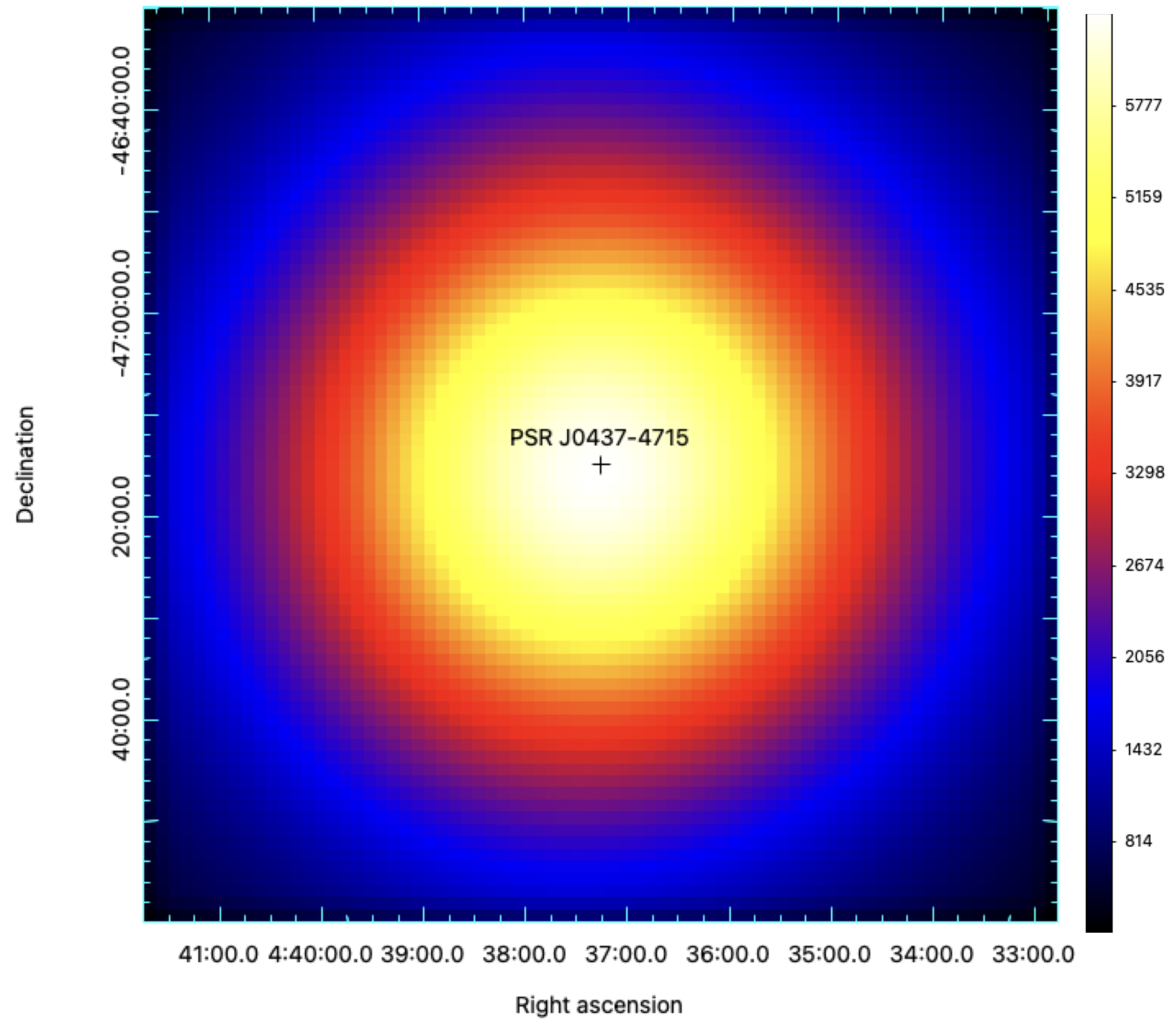}
    \includegraphics[width=0.74\linewidth]{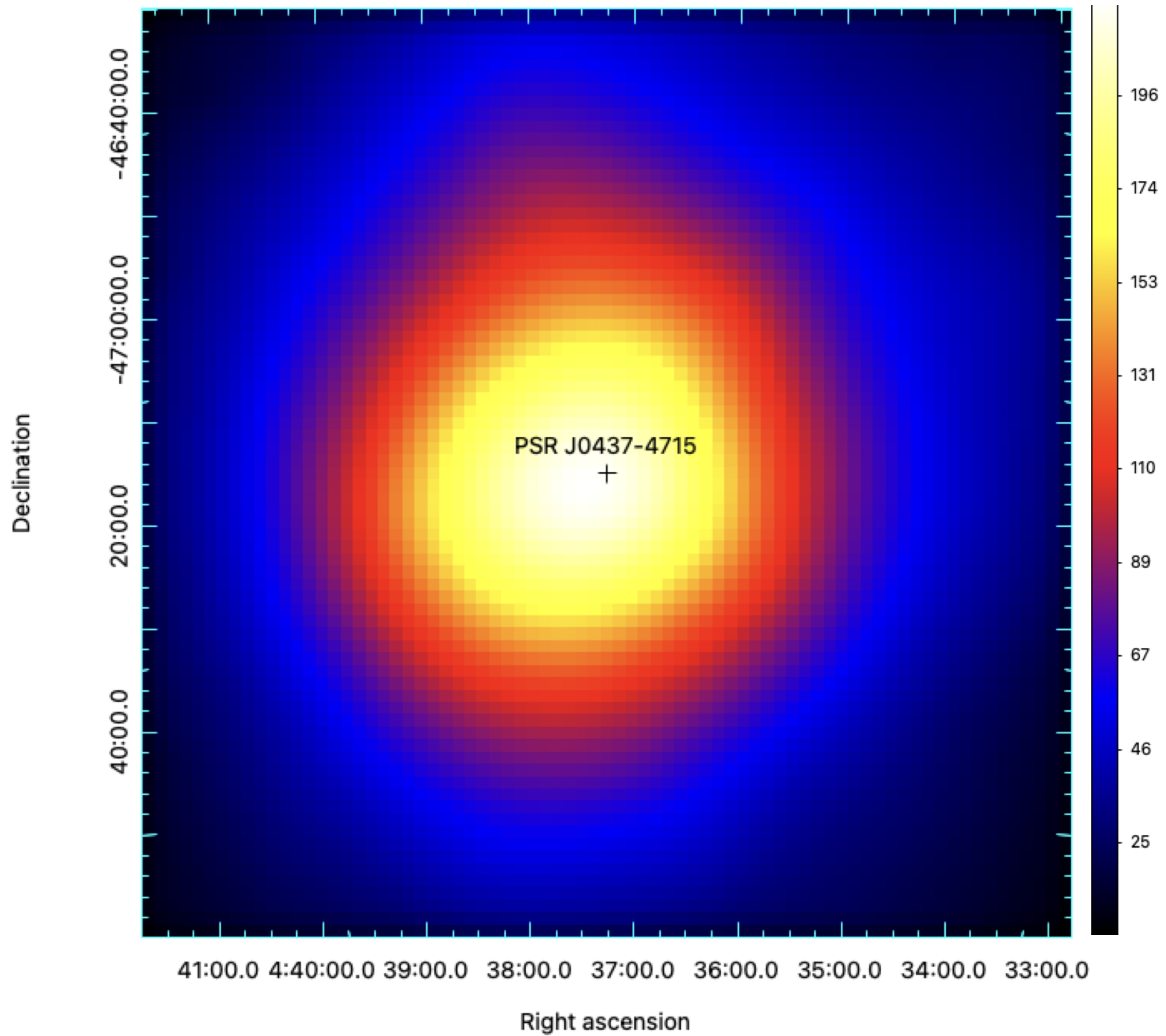}
    \caption{\g\ TS map (300 MeV to 300 GeV) of PSR~J0437$-$4715 from the whole data (top) and the off-pulse data (bottom).}
    \label{fig:ts-map}
\end{figure*}

PSR~J0437$-$4715 is bright in the \lat\ energy band and located in a clean field. The $1.5^{\circ} \times 1.5^{\circ}$ test statistic (TS) map were calculated for the source region from the whole LAT data and the off-pulse data, which were presented in the Figure~\ref{fig:ts-map}. The test statistic (TS) is adopted to estimate the significance of \g\ sources. It is defined by TS = 2 (ln $L_1$ - ln $L_0$), where $L_1$ and $L_0$ represent maximum likelihood values for background with target source and without target source. We conduct the analysis by searching radius of PSR~J0437$-$4715 in the off-pulse phase. Extension is an important factor in determining the nature of \g\ sources. A source extension analysis is executed by performing a likelihood ratio test with respect to the point source and the extended source: TS$_{\rm ext}$ = -2 $\left( \mathrm{ln}\left(L_{\rm PS}\right) - \mathrm{ln}\left(L_{\rm ext}\right) \right)$. A radial Gaussian model is used to obtain TS$_{\rm ext}$. The best-fit extension values are determined by performing a likelihood profile scan over the 68\% containment and fitting for the extension which maximizes the model. TS$_{\rm ext}$ for the off-pulse component is 5.9. So we obtain an upper limit on the radius of 0.12$^{\circ}$.

\begin{figure*}
    \centering
    \includegraphics[width=0.9\linewidth]{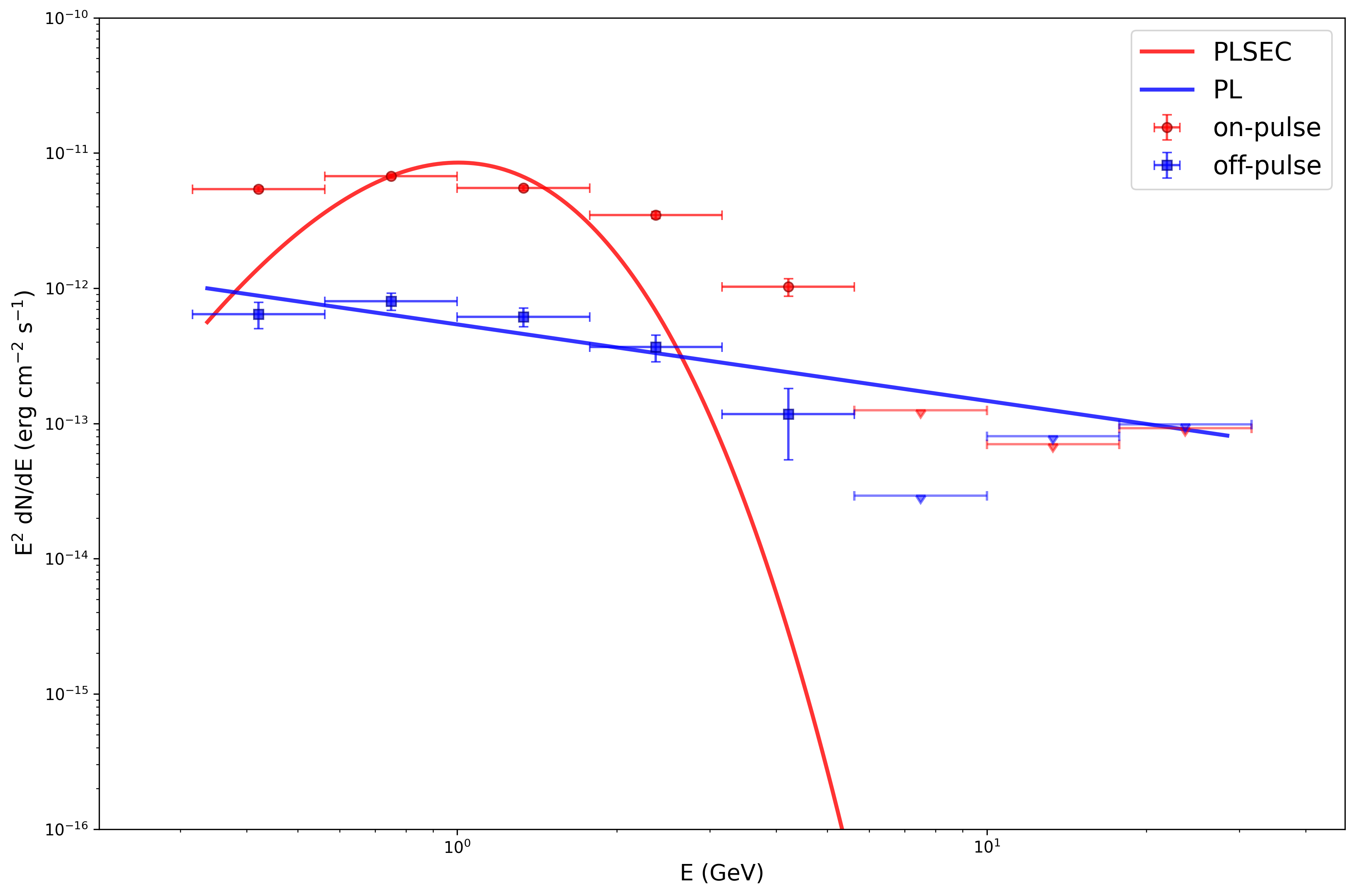}
    \caption{\g\ spectra (300 MeV to 300 GeV) of PSR~J0437$-$4715 in the on-pulse data and off-pulse data. The on-pulse data was modeled by a PLSC model while the off-pulse data was modeled by a PL model.}
    \label{fig:sed}
\end{figure*}

\subsection{Spectral Analysis}

The spectral shape aids us in judging the nature of \g\ sources. To validate the spectral model, we test for spectral curvature that reveals deviations from a PL spectrum for each source via likelihood ratio test. For a PLSC model, it gives: TS$_{\rm PLSC}$ = -2 $\left( \mathrm{ln}\left(L_{\rm PL}\right) - \mathrm{ln}\left(L_{\rm PLSC}\right) \right)$. Once TS$_{\rm PLSC}$ is larger than 25, we perform a spectral fit for PLSC models, otherwise, the PL model is adopted. The curvature test shows a marginal value TS$_{\rm PLSC} = 23$ for off-pulse. Therefore, a PLSC model should be considered also. Here we adopt PL as spectral model to fit the data of PSR~J0437$-$4715, which yields a spectral index $\Gamma = 2.71 \pm 0.09$ (Table~\ref{tab:fit-result}). Corresponding spectra can be found in Figure~\ref{fig:sed}. While for PLSC model with possible pulsar magnetospheric origin, we discuss in Section~\ref{sec:magnetosphere}.

\section{Understanding the Gamma-ray Emission of PSR~J0437$-$4715} \label{sec:orgin}

Off-pulse GeV emission of \g\ pulsars may help us to search for potential PWN, pulsar halo or bow-shock nebula around the pulsar \citep{Ackermann2011,Li2018,Zheng2023,Zheng2024,Lange2025}. In this work, we analyzed 17 years of \lat\ data for PSR~J0437$-$4715, which is the closest pulsar. To investigate the potential \g\ emission arising from outside the pulsar pulse, we perform analysis on off-pulse data. The detection of off-pulse emission from the MSP presents an intriguing scenario regarding the origin of its high-energy \g\ radiation. 

\begin{figure*}
    \centering
    \includegraphics[width=0.88\linewidth]{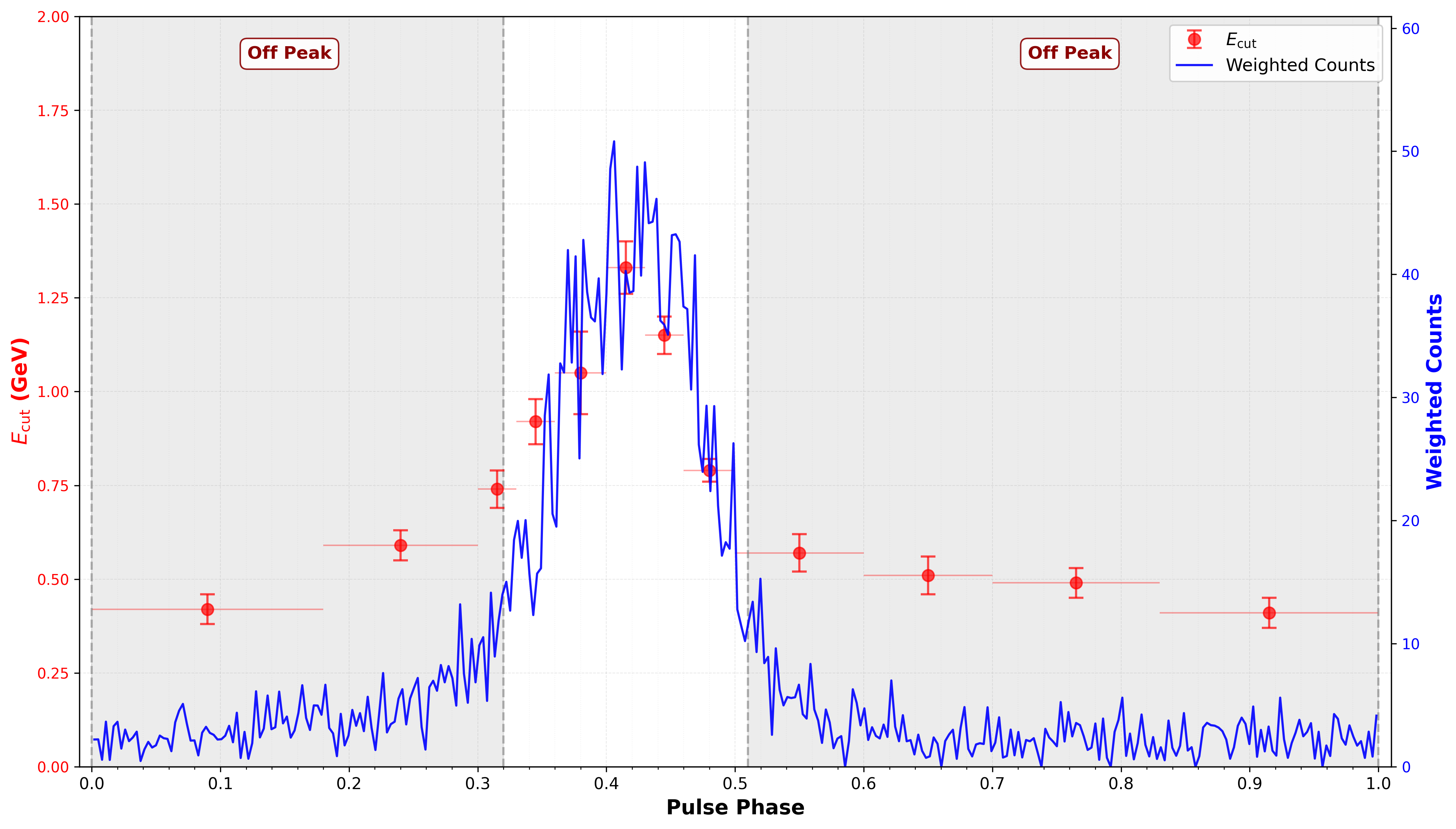}
    \caption{Spectral cutoff energy $E_{\rm cut}$ values (red points) as a function of pulse phase. The blue line show the weighted counts of PSR~J0437$-$4715. Off-pulse regions are shown as gray shadow.}
    \label{fig:phase-cutoff}
\end{figure*}

\begin{figure*}
    \centering
    \includegraphics[width=0.88\linewidth]{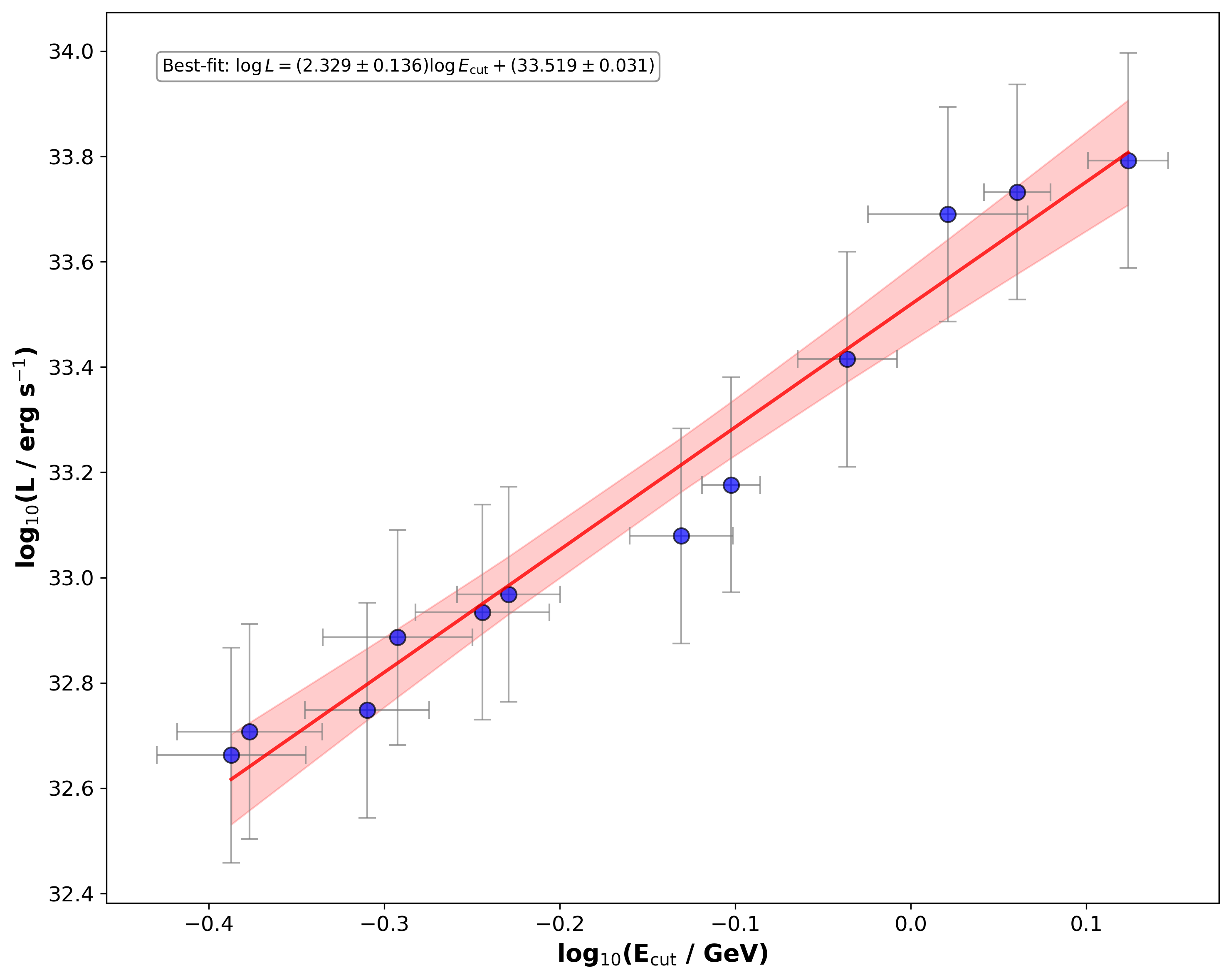}
    \caption{Linear correlation between \g\ luminosity $\log_{10}E_{\rm cut}$ and cutoff energy $\log_{10}L$ in different phase bin. The line and shadow represent the best-fit results with error. The spectral cutoff energy values are provided in Figure~\ref{fig:phase-cutoff}.}
    \label{fig:cut-lum}
\end{figure*}

\subsection{Magnetospheric Origin of Off-pulse Emission?} \label{sec:magnetosphere}

Since we obtain a marginal value of TS$_{\rm PLSC}$, we also consider the magnetospheric origin of off-pulse component. Thus, we investigate the whole phase data with another kinds of PL model with cutoff energy, which may help us to understand the current sheet of \g\ pulsars. This spectral model is defined by:

\begin{equation}
    \frac{dN}{dE} = N_0 (\frac{E}{E_0})^{-\Gamma}\exp [-(\frac{E}{E_{\rm cut}})^b]
\end{equation}

where $E_{\rm cut}$ is the cutoff energy, $b$ is the index describing the sharpness of the cutoff. We adopt $b=1$ as done by other work \citep{Lei2026}. Notice that it shares the same physical meaning with Equation~(\ref{eq:plsc}). A phase-resolved spectral analysis was conducted. Figure~\ref{fig:phase-cutoff} presents the best-fit cutoff energy in different phase bin of PSR~J0437-4715. It shows the variations of the spectral cutoff energy across the rotation phase.

Figure~\ref{fig:cut-lum} gives the relationship between $\log E_{\rm cut}$ and $\log L$ in different phase bin. Correlation is found between $\log E_{\rm cut}$ and $\log L$, which is $\log (L) = \alpha \log (E_{\rm cut}) + \beta$ with a best-fit $\alpha = 2.33 \pm 0.14$ and $\beta = 33.52 \pm 0.03$. This is similar to the value of equatorial current sheet (ECS) model \citep{Lei2026}. The consistency between the observed and predicted slopes provides strong observational evidence that the \g\ emission from MSPs is governed by curvature radiation in the ECS beyond the light cylinder, and establishes the $L_{\gamma}$ - $E_{\rm cut}$ relation as a key diagnostic for distinguishing emission mechanisms in pulsar magnetospheres.

\subsection{Conversion Efficiency and Its Implications for a Bow-shock PWN} \label{sec:efficiency}

Considering $L_{\gamma} = \eta \dot{E}$, where $\eta$ is conversion efficiency and $\dot{E}$ is the spin-down energy loss \citep{Smith2023}, we obtained $\eta = 0.004$. A \g\ pulsar can be explained by the outer gap model \citep{Cheng1986}. This model proposes that high-energy \g\ emission from a pulsar originates from a vacuum gap region in the outer magnetosphere, where particles are accelerated to relativistic energies by a strong electric field parallel to the magnetic field lines and subsequently emit curvature radiation \citep{Vigano2015,Takata2016}. To assess whether the low observed efficiency could be accommodated by a magnetospheric origin, we compare it with the theoretical prediction of the outer gap model. The theoretical conversion efficiency of MSP in the outer gap model was derived by \citep{Zhang2003}:

\begin{equation}
    \eta_{\rm th} = 4.7 \times 10^{-12}P^{20/7}\dot{P}^{-6/7}
\end{equation}

The derived $\eta_{\rm th} = 0.058$, which is higher than $\eta_{\rm obs} = 0.004$. This value is significantly lower than the typical total \g\ efficiency of young pulsars and lies at the lower end of the range observed for MSPs. Such an efficiency is lower than most of values predicted by the $\gamma$-ray emission with phase-locked to the pulsar spin. In these models, any off-pulse leakage would require extreme gap geometries or very large inclination angles, yet even then the predicted off-pulse fraction rarely exceeds a few percent of the on-pulse flux, and the spectral index tends to be harder ($\Gamma\lesssim 2.0$) due to curvature radiation. This spectral shape is naturally produced by synchrotron or inverse Compton (IC) emission from relativistic electrons accelerated at a collisionless shock, which is typical for a bow-shock PWN.

The measured $\eta \approx 0.4\%$ implies that only a small fraction of the pulsar's spin-down power is converted into relativistic electrons that subsequently radiate in the GeV band. This is consistent with a termination shock located very close to the pulsar, where the post-shock magnetic field is modest and IC cooling on the cosmic microwave background (CMB) dominates over synchrotron.

\subsection{Alternative Scenarios} \label{sec:scenarios}

Alternative scenarios that could in principle produce GeV emission include the pulsar's binary companion wind interaction \citep{Bednarek2013} and the pulsar's own halo via IC scattering of background photons \citep{Giacinti2020,Lopez2022}. However, the companion of PSR~J0437$-$4715 is a cool white dwarf with a weak outflow \citep{Verbiest2008}, making a shock between the pulsar wind and the companion wind too faint to account for the observed flux. Another possibility is that the off-pulse emission originates from a large-scale halo powered by escaping pairs, as seen for some middle-aged pulsars \citep{Abeysekara2017,Cao2025}. Our spatial analysis rules out extension beyond $0.12^{\circ}$, which disfavors a diffusion-dominated halo. Consequently, the bow-shock PWN of PSR~J0437$-$4715 is expected to be a faint, point-like GeV emitter.

\section{$\gamma$-ray From Bow-shock Pulsar Wind Nebula} \label{sec:population}

The above analyses reveal that the off-pulse emission of PSR~J0437$-$4715 is spectrally soft (
$\Gamma = 2.71 \pm 0.09$) and spatially unresolved (extension upper limit $< 0.12^{\circ}$), which are suggestive of either a compact bow-shock PWN or a magnetospheric leakage. However, single-source diagnostics alone cannot discriminate between these scenarios, because the predicted angular sizes of bow-shock nebulae for nearby MSPs often fall below the Fermi-LAT point-spread function, and the spectral curvature may be affected by phase-selection biases. To overcome this limitation and to place PSR~J0437$-$4715 in a broader physical context, we next examine a sample of 17 pulsars with confirmed bow-shock PWNe in X-rays \citep{Kargaltsev2017,Smith2023}. This comparative approach enables us to test whether the emission properties of PSR~J0437$-$4715 follow the same pattern that are expected from theoretical models of termination-shock acceleration. If the source aligns with the sample trends, its off-pulse emission is likely governed by the same shock-physics processes.

The bow-shock can produce synchrotron emission from electrons accelerated in relativistic collisionless shocks \citep{Bykov2017}. Consequently, the electron energy distribution can be described as a PL model \citep{Hededal2004}. Table~\ref{tab:bspwn} provides different parameters from \g\ pulsars with detected bow-shock in X-ray. Most of them (including bow-shock radius $R_{\rm bs}$ and X-ray luminosity $\log L_{\rm X}$) were summarized by \cite{Kargaltsev2017}, while \g\ luminosity were calculated by \cite{Smith2023}.

\begin{figure*}
    \centering
    \includegraphics[width=0.48\linewidth]{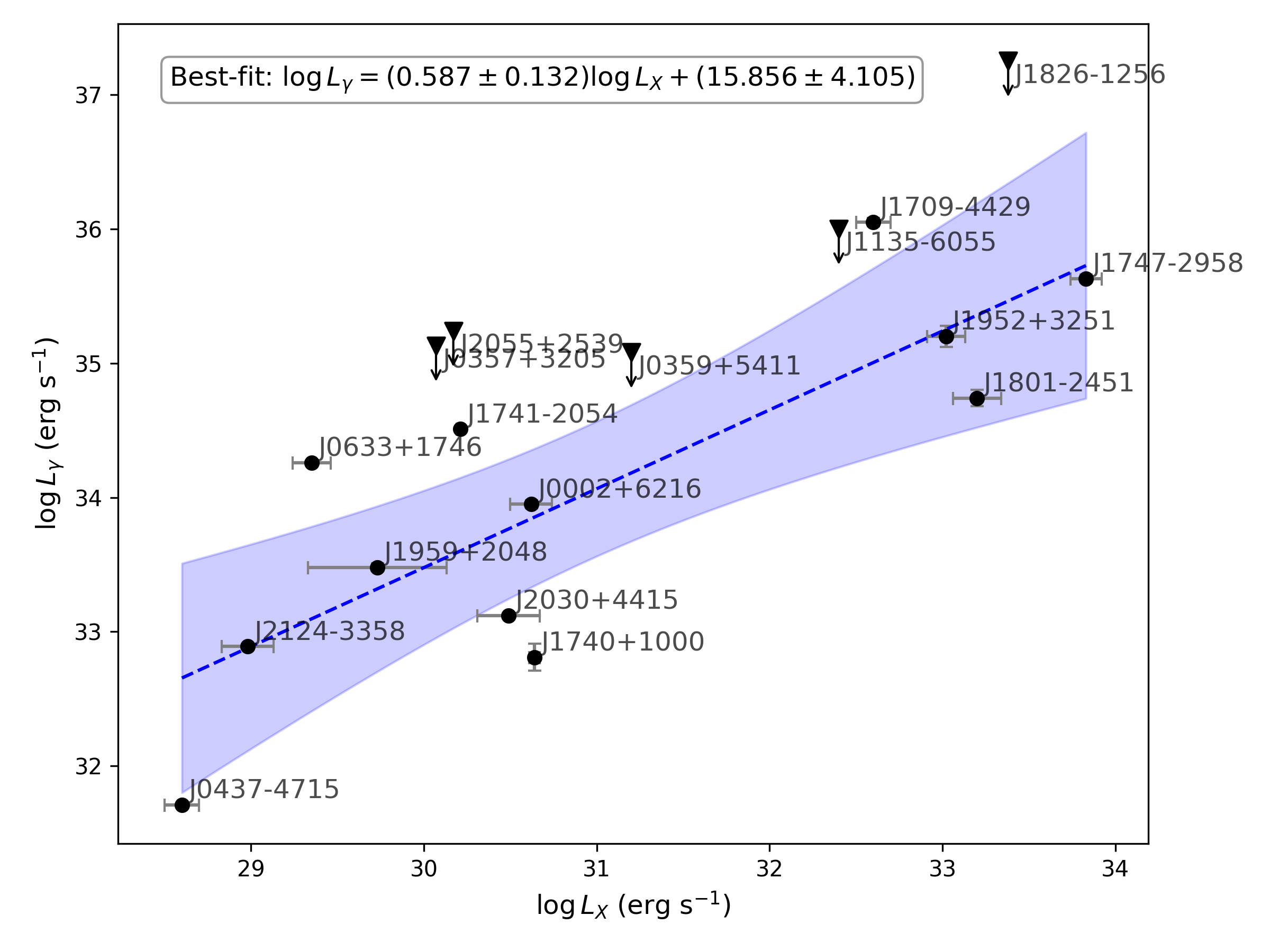}
    \includegraphics[width=0.48\linewidth]{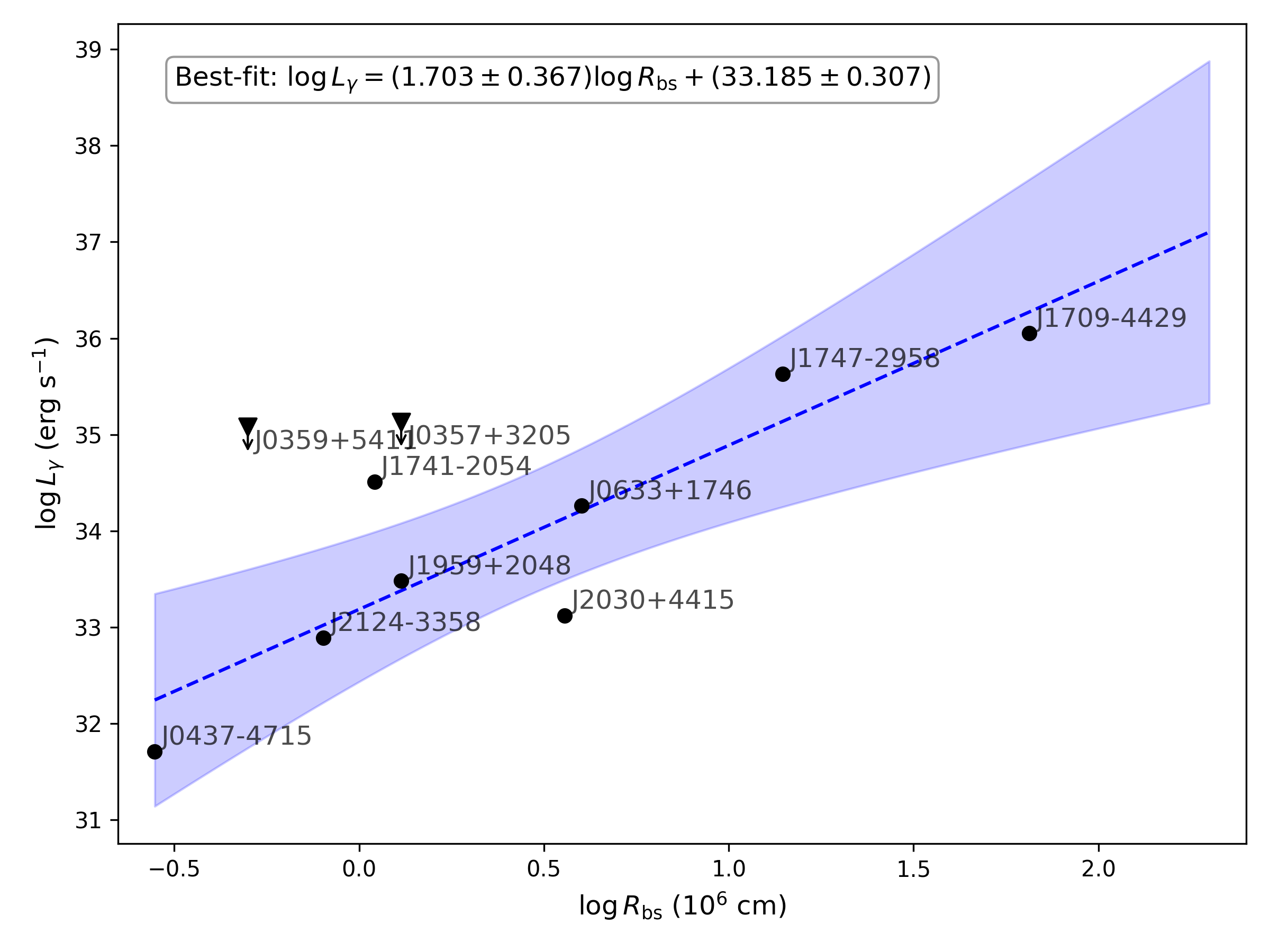}
    \caption{Left panel: \g\ luminosity $L_{\gamma}$ vs. X-ray luminosity $L_{X}$. Right panel: \g\ luminosity $L_{\gamma}$ vs. bow-shock radius $R_{\rm bs}$. The dashed lines and the gray shadows provide the best-fit results of each panel.}
    \label{fig:bow-shock-fig}
\end{figure*}

The correlation between \g\ luminosity $L_{\gamma}$ and X-ray luminosity $L_{X}$ in our bow-shock PWN sample provides critical insight into the origin of the \g\ emission and the dominant Compton scattering regime. A linear relation would suggest that the \g\ arise from IC scattering of external photon fields such as the CMB or interstellar radiation field, while the X-rays are produced via synchrotron radiation from the same relativistic electron population. The X-ray synchrotron emission and the \g\ emission are energetically coupled through the same electron spectrum, so that a change in the electron population would affect both X-ray and \g\ consistently. In this case, a near-constant ratio between the external radiation energy density and the magnetic field energy density should happen. Conversely, a relation $L_{\gamma} \sim L_{X}^2$ would strongly favor the synchrotron self-Compton (SSC) process, where the same electrons up-scatter their own synchrotron photons to \g\ energies. Such a relation would indicate that the \g\ and X-ray emitting regions are co-spatial and that the magnetic field is relatively weak. Figure~\ref{fig:bow-shock-fig} (left panel) presents the relationship between $L_{\gamma}$ and $L_{X}$. The best-fit relation yields $L_{\gamma} \sim L_{X}^{0.61 \pm 0.21}$. To physically interpret this scaling, we relate the observed luminosities to the underlying relativistic electron population. Assuming a power‑law electron energy distribution $N_e \propto \gamma^{-p}$ accelerated at the termination shock, the synchrotron X‑ray luminosity scales as $L_{X} \propto B^2 \gamma_{\rm max}^{3-p}$. For IC scattering of external soft photons in the Thomson regime, the GeV luminosity follows $L_{\gamma} \propto U_{\rm ph} \gamma_{\rm max}^{3-p}$, where $U_{\rm ph}$ energy density of photon fields. If the same electron population produces both emissions and the ratio $U_{\rm ph}/B^2$ is roughly constant across the sample, a linear correlation $L_{\gamma} \propto L_X$ is expected. The observed slope of 0.61, however, is significantly shallower than unity. This deviation indicates that the magnetic field strength B is not universal among bow‑shock nebulae; sources with higher X‑ray luminosity likely possess stronger post‑shock magnetic fields, which enhances synchrotron cooling and suppresses the high‑energy tail of the electron spectrum. As a result, the GeV IC emission does not increase proportionally with the X‑ray emission, leading to a sub‑linear scaling.

The derived relation has several important implications. First, it firmly rules out the synchrotron self‑Compton (SSC) process as the dominant mechanism for the observed GeV emission, because SSC would require $L_{\gamma} \propto L_X^2$, since the upscattered photon density scales with the synchrotron energy density. The measured slope is away from 2, providing evidence that the \g\ are predominantly produced by IC scattering of external radiation fields rather than by SSC. Second, the sub‑linear slope implies that the magnetic energy density varies considerably among the sample; a constant magnetic field would have preserved a linear relation. The observed trend suggests that bow‑shock nebulae with larger X‑ray luminosity tend to have stronger magnetic fields, which preferentially cool the highest‑energy electrons and thus reduce the efficiency of GeV emission. Finally, we caution that the sample may include pulsars at different evolutionary stages or with different ambient densities, which could also contribute to the scatter.

The relationship between \g\ luminosity  $L_{\gamma}$ and bow-shock radius $R_{\rm bs}$ serves as a powerful diagnostic for the energy source of relativistic particles and the efficiency of particle acceleration at the termination shock. The bow-shock radius is determined by the pressure balance between the pulsar wind and the surrounding ISM, scaling as: 

\begin{equation}
    R_{\rm bs} = \frac{\dot{E}}{4 \pi c \rho_{\rm ISM}v_{\rm psr}^2}
\end{equation}

where $\dot{E}$ is the spin-down power of the pulsar, $\rho_{\rm ISM}$ is the density of ISM, and $v_{\rm psr}$ is the velocity of pulsar. This relation is developed by \cite{Romani2022}. A scaling of $L_{\gamma} \sim R_{\rm bs}^2$ indicates that the \g\ luminosity is proportional to the spin-down power, suggesting a roughly constant efficiency in converting the pulsar wind kinetic energy into \g\ emitting electrons. Figure~\ref{fig:bow-shock-fig} (right panel) provides a plot of $L_{\gamma}$ and $R_{\rm bs}$, which gives the derived relation: $L_{\gamma} \sim R_{\rm bs}^{1.70\pm0.73}$. A scaling
$L_{\gamma} \sim R_{\rm bs}^{\alpha}$ with $\alpha > 2$ would point to an enhanced acceleration efficiency or reduced cooling losses in larger bow shocks. However, \g\ emission dominated by particles accelerated from the ISM can not be ruled out. The \g\ luminosity is probably also governed by other factors, such as the magnetic field strength, the injection spectral index, or the viewing geometry.

The best‑fit scaling $L_{\gamma} \sim R_{\rm bs}^{1.70\pm0.73}$ suffers from a large statistical uncertainty. Consequently, the derived slope is compatible with both $\alpha = 1$ and$\alpha=2$. Therefore, our data cannot distinguish between a constant conversion efficiency (which would predict $\alpha=2$, as $R_{\rm bs}^2 \propto \dot{E}$ under constant ISM density) and other dependencies. Moreover, the significant scatter seen in Figure~\ref{fig:bow-shock-fig} (right) – for instance, PSR~J0633$+$1746 and PSR~J0437$‑$4715 have similar bow‑shock radii but differ markedly in $L_{\gamma}$ – indicates that factors other than the shock radius, such as the local magnetic field strength, the inhomogeneity of the ambient medium, and the projection effect of the tail orientation, can substantially modulate the observed GeV luminosity. We thus conclude that the current sample does not provide robust evidence for a universal efficiency of converting spin‑down power into GeV emission; instead, the $L_{\gamma} - R_{\rm bs}$ relation is weakly constrained and should be revisited with a larger and more homogeneously measured sample, preferably including direct measurements of the interstellar density and the pulsar's proper motion angle.

\begin{table*}
    \centering
    \begin{tabular}{l|c|c|c|c|c|c|c|c}
    \hline
    PSR &  $d$	& log$\dot{E}$& log $\tau$	& $B_{11}$     & $v_{\perp}$	& $r_{\rm bs}$ & log $L_{\rm X}$ &   log $L_{\gamma}$\\
        &  kpc  & erg/s      &    yr        &  $10^{11}$ G & km/s & $10^6$ cm  & erg/s  &   erg/s \\
    \hline
    J1952$+$3251	& 	3         & 36.57	& 5.03	& 4.86	& 460          & 	$<12$	& $33.02\pm0.11$  & $35.20 \pm 0.08$\\
    J1826$-$1256	&	3.9     & 36.56	& 4.16	& 37	&... 	       &...	& $33.38\pm0.06$  & $<37.25$\\
    J1709$-$4429	&	2.6	      & 36.53	& 4.24	& 31.2	&$\lesssim100$ &	$\sim70$	&$32.60\pm0.10$	& $36.05 \pm 0.03$\\
    J1801$-$2451  &	3.8	      & 36.41   & 4.19	& 40.4	&198 	       &$<13$	& $33.20\pm0.14$ & $34.74 \pm 0.06$\\
    J1747$-$2958	&	5	      & 36.4    & 4.41  & 24.9  &	$306\pm43$ &	$\sim14$&	$33.83\pm0.09$ & $35.63 \pm 0.03$\\
    J1135$-$6055	&	2.8      &	36.32   & 4.36  & 30.5  &	$<330$     &	...	&$32.40\pm0.04$	 &  $<36.00$ \\
    J1959$+$2048	&	1.73	  &35.2	    & 9.18	& 0.002	&$\sim220$	   & 1.3	& $29.73\pm0.40$ & $33.48 \pm 0.03$\\
    J0359$+$5411	&  1.04       &	34.66   & 5.75	& 8.39	&$61^{+12}_{-9}$&	0.5	& $31.20\pm0.07$  & $<35.08$\\
    J0633$+$1746	&	0.25      &	34.51   & 5.53	& 16.3	&$\sim$200 	   & 4&	$29.35\pm0.11$ &  $34.26 \pm 0.01$\\
    J2030$+$4415	&	1       &34.46	& 5.74	& 12.3	&... 	       & 3.6	&$30.49\pm0.18$	 & $33.12 \pm 0.02$ \\
    J1741$-$2054	&	0.3	      &33.97	& 5.59	& 26.8	&155           &	1.1	& $30.21\pm0.02$	 & $33.11 \pm 0.01$\\
    J2124$-$3358	&	0.41	  &33.83	& 9.58	& 0.003	& 75           &	0.8	& $28.98\pm0.15$	 & $32.89 \pm 0.01$\\
    J0357$+$3205	&	0.5	      & 33.77	& 5.73	& 24.3  & $\sim$2000 &1.3	&$30.07\pm0.20$	 & $<35.13$\\
    J0437$-$4715	&	0.156     &	33.74	& 9.2	& 0.006	& $104.7\pm0.9$& 0.28 &$\sim$28.6	 & $31.71 \pm 0.01$\\
    J2055$+$2539	&	0.6     &	33.69   & 6.09	& 11.6  &	$\lesssim2300$ &	$<1.4$	&$30.17\pm0.03$	& $<35.24$\\
    J1740$+$1000	&	1.22       &  35.36      &	5.06      & 18.3	  &	$\sim 2800$ & ...   &	$30.64 \pm 0.03$  &	 		$32.81 \pm 0.10$	\\
    J0002$+$6216	&	6.36       &   35.18      &	  5.49    &	8.4     & ...  &  ...   & $30.62 \pm 0.12$ &	$33.95 \pm 0.03$ \\
    \hline
    \end{tabular}
    \caption{Columns give: pulsar name, distance, energy loss rate, characteristic age, surface magnetic field strength $B_{11}$, projected velocity, bow-shock radius, X-ray luminosity, and \g\ luminosity. The reference of most of them can be found in Table 1 and Table 2 of \citet{Kargaltsev2017}. We add PSR J1740$+$1000 and PSR J0002$+$6216, which are pulsars with bow-shock. All \g\ data are adopted from \citet{Smith2023} (3PC).}
    \label{tab:bspwn}
\end{table*}

\section{Conclusions}

The pulsar wind shock within the bow shock of the binary systems may produce \g\ emissions. The observed H$\alpha$ and non-thermal X-ray emissions were observed in PSR~J0437$-$4715, which indicate relativistic electrons acceleration \citep{Brownsberger2014,Rangelov2016}. The inverse Compton scattering may happen between electron accelerated by pulsar wind and background photon including the CMB and the diffuse optical and infrared background. This scenario predicts extended \g\ emissions around the binary system. However, no extended emission was found in PSR~J0437$-$4715.

In older MSP systems, the wind nebula can be compressed by the ambient medium or the binary companion's outflows, resulting in a scale that is unresolved by the \lat\ . Our non-detection of extended emission in the off-pulse phase of PSR~J0437$-$4715 implies that either the putative bow shock PWN is too compact to be resolved by \lat\ , or its inverse Compton emission is too weak compared to the diffuse background. Crucially, when we place PSR~J0437$-$4715 into the larger sample of 17 bow-shock systems, we find that its observed $L_{\gamma}$, $L_X$ and $R_{\rm bs}$ are consistent with the global scaling relations derived from the sample . This supports the interpretation that the off-pulse GeV emission from PSR~J0437$-$4715 shares a common physical origin with those of other bow-shock pulsars: IC scattering of external photons by relativistic electrons accelerated at the termination shock rather than being a magnetospheric leakage or a binary-wind interaction product. The combined single source and population study thus strengthens the case that bow-shock PWN are intrinsically faint, compact GeV emitters, and that their radiative properties are primarily governed by the pressure balance between the pulsar wind and the ambient ISM. Nevertheless, the large scatter in the $L_{\gamma}-R_{\rm bs}$ relation and the sub-linear slope of $L_{\gamma}-L_X$ indicate that secondary factors, such as inhomogeneous ISM density, may play a non-negligible role. This motivates deeper observations with future telescopes such as Cherenkov Telescope Array \citep{cta2019}, which may resolve such structures and constrain the contribution of MSP nebulae to the cosmic ray positron flux.

Statistical relations from a bow-shock PWN sample imply $L_{\gamma} \sim L_{X}^{0.61 \pm 0.21}$ and $L_{\gamma} \sim R_{\rm bs}^{1.70 \pm 0.73}$. The former suggests that IC scattering of external photon fields likely dominates the \g\ emission, although SSC can not be ruled out. The later indicates that \g\ luminosity scales with bow-shock radius, consistent with a roughly constant efficiency of converting pulsar wind energy into GeV-emitting electrons.

\section*{Acknowledgements}

This work is supported by the National Natural Science Foundation of China (NSFC, Grant No. 12393853). The author appreciates Professor Hao Zhou’s help, as well as the support from Tsung-Dao Lee Institute. 

\section*{Data Availability}

The \lat\ data used in this work are publicly available and are provided online at the NASA-GSFC Fermi Science Support Center \footnote{https://fermi.gsfc.nasa.gov/cgi-bin/ssc/LAT/LATDataQuery.cgi}.



\bibliographystyle{mnras}
\bibliography{example} 


\bsp	
\label{lastpage}
\end{document}